\documentclass[12pt]{article}

\usepackage[margin=1in]{geometry}
\usepackage[onehalfspacing]{setspace}
\usepackage[T1]{fontenc}
\usepackage[utf8]{inputenc}
\usepackage[numbers]{natbib}
\usepackage{amsmath, amssymb, amsthm, mathtools}
\usepackage{bm}
\usepackage{booktabs}
\usepackage{array}
\usepackage{graphicx}
\usepackage{subcaption}
\usepackage{float}
\usepackage{url}
\usepackage{hyperref}
\hypersetup{colorlinks=true, citecolor=blue, linkcolor=blue, urlcolor=blue}
\usepackage{authblk}
\usepackage{pdflscape}

\title{\vspace{-1.2in}
Robust Survival Estimation under Interval Censoring: Expectation–Maximization and Bayesian Accelerated Failure Time Assessment via Simulation and Application}
\author[1]{Joshua Korley}
\affil[1]{Arnold School of Public Health, University of South Carolina, Columbia, SC, USA}
\date{}

\begin{document}
\maketitle

\begin{abstract}
Interval censoring occurs when event times are only known to fall between scheduled assessments, a common design in clinical trials, epidemiology, and reliability studies. Standard right-censoring methods, such as Kaplan--Meier and Cox regression, are not directly applicable and can produce biased results. This study compares three complementary approaches for interval-censored survival data. First, the Turnbull nonparametric maximum likelihood estimator (NPMLE) via the EM algorithm recovers the survival distribution without strong assumptions. Second, Weibull and log-normal accelerated failure time (AFT) models with interval likelihoods provide smooth, covariate-adjusted survival curves and interpretable time-ratio effects. Third, Bayesian AFT models extend these tools by quantifying posterior uncertainty, incorporating prior information, and enabling interval-aware model comparisons via PSIS--LOO cross-validation. Simulations across generating distributions, censoring intensities, sample sizes, and covariate structures evaluated the integrated squared error (ISE) for curve recovery, integrated Brier score (IBS) for prediction, and coverage for uncertainty calibration. Results show that the EM achieves the lowest ISE for distribution recovery, AFT models improve predictive performance when families are correctly specified, and Bayesian AFT offers calibrated uncertainty and principled model selection. An application to the ovarian cancer dataset, restructured into interval-censored form, demonstrates the workflow in practice: the EM algorithm reveals the baseline shape, parametric AFT provides covariate-adjusted predictions, and Bayesian AFT validates model adequacy through posterior predictive checks. Together, these methods form a tiered strategy: EM for shape discovery, AFT for covariate-driven prediction, and Bayesian AFT for complete uncertainty quantification and model comparison.
\medskip

\end{abstract}

\section{Introduction}
In many biomedical, engineering, and reliability studies, the event of interest is known to occur only between two scheduled assessments. This phenomenon, known as \emph{interval censoring}, differs fundamentally from right censoring because the event time $T_i$ for subject $i$ is constrained to lie within an interval $(L_i, R_i]$ rather than being known only to exceed a single threshold \citep{Turnbull1976, Sun2006}. Standard tools designed for right-censored data, such as the Kaplan--Meier estimator \citep{KaplanMeier1958} or the Cox proportional hazards model \citep{Cox1972}, are not directly applicable in this setting and can yield biased estimates of survival functions or regression effects when naively applied to interval-censored data.

A major breakthrough was achieved by Turnbull \citep{Turnbull1976}, who derived the nonparametric maximum likelihood estimator (NPMLE) under arbitrary interval-censoring. This estimator, obtained via an expectation--maximization (EM) procedure, allocates mass to a set of disjoint intervals, commonly referred to as ``Turnbull intervals''and is fully data-driven without imposing structural assumptions on the hazard. The underlying EM updates follow the self-consistency principle introduced by Efron \citep{Efron1967}. Subsequent methodological developments established efficient computation using iterative convex minorant (ICM) algorithms and convex optimization theory \citep{GroeneboomWellner1992, Jongbloed1998} and regression extensions under proportional hazards assumptions \citep{Finkelstein1986}. The NPMLE remains the foundation of nonparametric interval-censored survival analysis.

Parametric approaches, such as accelerated failure time (AFT) models with Weibull or log-normal distributions, yield smooth curves, admit covariates, and provide interpretable regression effects \citep{KalbfleischPrentice2002, KleinMoeschberger2003}. In particular, the coefficients in the AFT models correspond to multiplicative shifts in the median or mean survival times. These advantages make AFT models attractive for predictions and extrapolations. However, their limitation lies in distributional misspecification: when the true data-generating process deviates from the assumed family, especially in the tails or under heavy censoring, the resulting estimates may be biased.

Bayesian survival models extend parametric AFT formulations by placing priors on regression coefficients and shape parameters, thereby enabling full posterior inference and principled uncertainty quantification \citep{IbrahimChenSinha2001, Buerkner2017}. The likelihood contribution for interval-censored observations is naturally expressed as the survival probability mass between $L_i$ and $R_i$, which integrates seamlessly into the Bayesian framework. Posterior predictive checks (PPCs) provide model-based diagnostics by assessing whether replicated data from the fitted posterior allocate the appropriate probability mass to each censoring interval \citep{Gelman1996}. Model comparison is facilitated by information criteria such as WAIC or PSIS--LOO cross-validation \citep{Vehtari2017, Carpenter2017}, offering an interval-aware approach to assessing the adequacy of competing families. With modern Hamiltonian Monte Carlo implementations, Bayesian AFT models are computationally tractable, even for moderately large datasets.

Interval-censored data arise frequently in biomedical research, for example, periodic imaging in oncology trials, incubation periods in infectious disease studies such as HIV/AIDS, and scheduled follow-ups in Alzheimer’s disease cohorts \citep{Sun2006, LindseyRyan1998, Zhang2009}. In engineering and reliability contexts, lifetime testing under inspection regimes also produces interval-censored outcomes \citep{Lawless2002}. These applications illustrate the necessity of methods that respect the interval structure; naive approaches such as midpoint imputation or treating right endpoints as exact event times can lead to severe bias in both survival estimation and covariate effects \citep{WellnerZhan1997}.

From a methodological perspective, interval censoring occupies an intermediate position between exact survival data and grouped outcome data. Unlike right censoring, which admits partial likelihood methods \citep{Cox1972, AndersenGill1982}, the interval-censored likelihood requires the integration of the density over subject-specific observation windows. This challenge has motivated developments in EM algorithms \citep{Turnbull1976}, self-consistency iterations \citep{Efron1967}, convex optimization approaches \citep{GroeneboomWellner1992}, and spline-based semiparametric models \citep{ZengLin2007, Sun2006}. Recent extensions include penalized regression, frailty models, and high-dimensional variable selection for complex interval-censored design.

\paragraph{Contributions.}
This study provides a unified comparative study of three complementary approaches for interval-censored survival analysis:  
(i) a derivation and implementation of the EM algorithm for the Turnbull NPMLE, serving as a nonparametric benchmark;  
(ii) Parametric Weibull and log-normal accelerated AFT models with interval likelihoods, incorporating covariates to yield interpretable time-ratio effects.  
(iii) Bayesian AFT models with weakly informative priors, enabling full uncertainty quantification, posterior predictive validation, and principled model comparisons.  

We evaluated these methods through simulation scenarios that varied the generating distribution, censoring intensity, sample size, and covariate structure, using integrated squared error (ISE), integrated Brier score (IBS), and empirical coverage as performance criteria. In addition, we applied the workflow to an ovarian cancer dataset, restructured into an interval-censored form by imposing periodic assessment windows, thereby demonstrating how the methodology extends to real clinical data.

\paragraph{Outline.}
Section~\ref{sec:setup} formalizes the data structure and likelihood of interval censoring. Section~\ref{sec:methods} presents the EM algorithm for the NPMLE, parametric and Bayesian AFT formulations, and the posterior computation. Section~\ref{sec:design} describes the simulation scenarios and the applied analysis of the ovarian dataset. Section~\ref{sec:results} reports the simulation and applied results, and Section ~\ref{sec:metrics} introduces the performance metrics (ISE, IBS, and coverage). Section~\ref{sec:discussion} discusses the implications and limitations of this study and suggests future research directions.

\section{Data structure and likelihood}\label{sec:setup}
Let $T_i$ denote the event time for subject $i=1,\dots,n$, with distribution function 
$F(t)=P(T_i \le t)$, survival function $S(t)=1-F(t)$, and density $f(t)=F'(t)$. Here, $f(t) = F'(t)$ denotes the event time density. 
Under interval censoring, we observe $(L_i, R_i]$ such that $P(L_i < T_i \le R_i \mid F) = 1$, 
where $0 \le L_i < R_i \le \infty$.

For interval-censored observations, the probability that $T_i$ lies within the observed interval $(L_i,R_i]$ can be expressed equivalently as
\[
P(L_i < T_i \le R_i \mid F) = F(R_i) - F(L_i) = S(L_i) - S(R_i),
\]
This links the distribution function $F(\cdot)$ and the survival function $S(\cdot)$.

Special cases include left-censoring ($L_i=0$), right-censoring ($R_i=\infty$), 
and exact observation ($L_i=R_i$). For exact events, the likelihood contribution 
uses $f(t_i;\theta)$. For observation $i$ the contribution to the likelihood is
\[
\mathcal{L}_i(\theta)=
\begin{cases}
S(L_i;\theta) - S(R_i;\theta), & 0<L_i<R_i<\infty \quad \text{(interval)},\\[4pt]
1 - S(R_i;\theta), & L_i=0,\, R_i<\infty \quad \text{(left)},\\[4pt]
S(L_i;\theta), & 0<L_i<\infty,\, R_i=\infty \quad \text{(right)},\\[4pt]
f(t_i;\theta), & L_i=R_i=t_i \quad \text{(exact)}.
\end{cases}
\]
Assuming independence, the full likelihood is
\[
L(\theta) = \prod_{i=1}^n \mathcal{L}_i(\theta), \qquad 
\ell(\theta) = \sum_{i=1}^n \log \mathcal{L}_i(\theta).
\]
This formulation follows the foundational contributions of Turnbull~\citep{Turnbull1976}, 
Sun~\citep{Sun2006}, Efron~\citep{Efron1967}, and Finkelstein~\citep{Finkelstein1986}.
 The form of censoring varies according to the design. Imaging studies and periodic clinic visits yield interval endpoints; administrative end and dropout induce right censoring; and left censoring arises when events precede the first visit. The unified likelihood above accommodates all cases and underpins both non-parametric and model-based estimations in our workflow.

\begin{figure}[H]
\centering
\includegraphics[width=.8\textwidth]{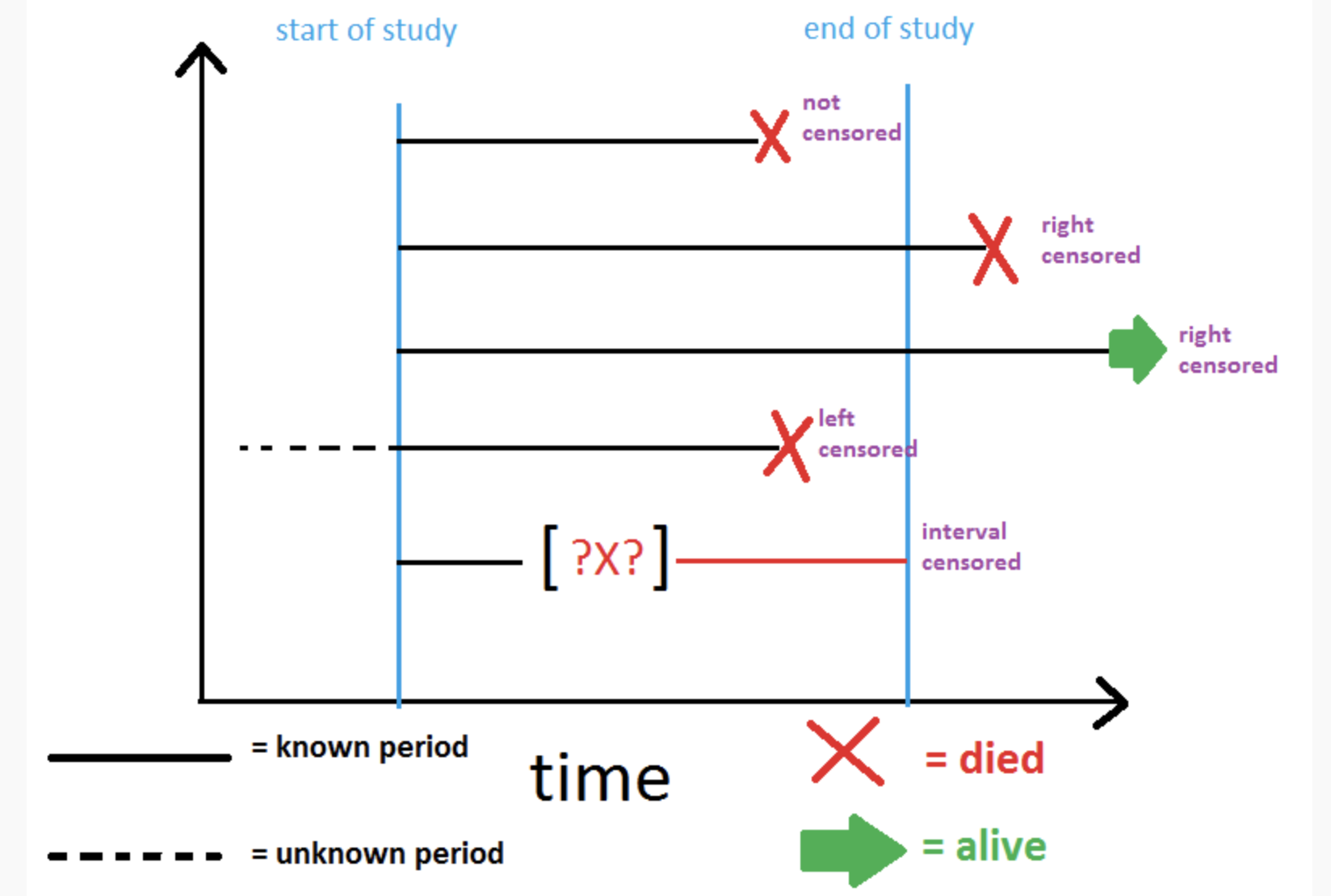}
\caption{Illustration of censoring types in survival analysis. Each horizontal line represents the follow-up of a participant. Solid segments indicate the observed status, and dashed segments indicate the unknown status. Red crosses mark events; green arrows indicate alive at last contact.}

\label{fig:interval}
\end{figure}
This unified likelihood forms the basis for both the nonparametric EM estimator and the parametric/Bayesian models described in Section~\ref{sec:methods}.
\section{Methods}\label{sec:methods}

\subsection{Nonparametric EM for the Turnbull NPMLE}
Suppose subject $i=1,\dots,n$ has true event time $T_i$ with survival function $S(t)=P(T_i>t)$, distribution function $F(t)=1-S(t)$, and density $f(t)=F'(t)$.  
Under interval censoring, we do not observe $T_i$ exactly, but only that it lies within an interval $(L_i,R_i]$, where $0 \le L_i < R_i \le \infty$.  

The observed likelihood contribution is therefore
\[
\mathcal{L}_i = P(L_i < T_i \le R_i) = F(R_i) - F(L_i) = S(L_i) - S(R_i).
\]
Special cases include left censoring ($L_i=0$), right censoring ($R_i=\infty$), and exact observation ($L_i=R_i$).  
Let the union of all observed endpoints define $m$ disjoint Turnbull intervals $\{I_j=(a_{j-1},a_j]\}_{j=1}^m$.  
The nonparametric maximum likelihood estimator (NPMLE) assigns masses $\bm{p}=(p_1,\dots,p_m)$ to these intervals, with $p_j\ge 0$ and $\sum_j p_j=1$.  
For subject $i$, let $\mathcal{J}_i=\{j:I_j \subseteq (L_i,R_i]\}$. Then
\[
L(\bm{p})=\prod_{i=1}^n \left(\sum_{j \in \mathcal{J}_i} p_j \right).
\]

Introducing latent variables $Z_i \in \{1,\dots,m\}$ to indicate the unobserved failure interval, the EM algorithm alternates as follows:
\[
\text{E-step: } w_{ij}^{(k)} = P(Z_i=j \mid \text{data}, \bm{p}^{(k)})
= \frac{p_j^{(k)}}{\sum_{r \in \mathcal{J}_i} p_r^{(k)}}, \quad j \in \mathcal{J}_i,
\]
\[
\text{M-step: } p_j^{(k+1)} = \frac{1}{n}\sum_{i=1}^n w_{ij}^{(k)}.
\]
This classical self-consistency update \citep{Turnbull1976, GroeneboomWellner1992} yields the Turnbull estimator, a step-function estimate of $F$ and hence $S$.  
It makes no distributional assumptions and provides a robust benchmark for subsequent parametric modeling.

\subsection{Parametric accelerated failure time (AFT) models}
Let $\bm{x}_i$ denote the vector of covariates for subject $i$.  
Under an accelerated failure time (AFT) model, the log survival time is modeled as
\[
\log T_i = \mu + \bm{x}_i^\top \bm{\beta} + \sigma \varepsilon_i,
\]
where $\mu$ is the location parameter, $\bm{\beta}$ is the regression coefficient, $\sigma$ is the scale parameter, and $\varepsilon_i$ follows a known distribution (e.g., extreme value for Weibull, normal for log-normal).  

For the Weibull AFT model, with shape parameter $\kappa>0$, the survival function is
\[
S(t \mid \bm{x}_i) = \exp\!\left[ - \left( \frac{t}{\exp(\mu+\bm{x}_i^\top \bm{\beta})}\right)^\kappa \right].
\]
The contribution of interval $(L_i,R_i]$ is then
\[
\mathcal{L}_i(\mu,\bm{\beta},\kappa) = S(L_i \mid \bm{x}_i) - S(R_i \mid \bm{x}_i).
\]
The coefficient interpretation is in terms of time ratios: $\exp(\beta_j)$ multiplies the median survival time for a one-unit increase in $x_j$.

\subsection{Bayesian accelerated failure time (AFT) models}
We extend the Weibull AFT to a Bayesian framework by combining the interval likelihood with priors.  
For subject $i$,
\[
\mathcal{L}_i(\mu,\bm{\beta},\kappa) = S(L_i \mid \bm{x}_i; \mu,\bm{\beta},\kappa) - S(R_i \mid \bm{x}_i; \mu,\bm{\beta},\kappa).
\]

The full likelihood is
\[
L(\mu,\bm{\beta},\kappa) = \prod_{i=1}^n \mathcal{L}_i(\mu,\bm{\beta},\kappa).
\]

We specify weakly informative priors:
\[
\mu \sim N(0, \sigma_\mu^2), \qquad 
\bm{\beta} \sim N(\bm{0}, \sigma_\beta^2 I), \qquad 
\kappa \sim \mathrm{Gamma}(a_\kappa,b_\kappa).
\]
The posterior distribution is then
\[
p(\mu,\bm{\beta},\kappa \mid \mathcal{D}) \;\propto\;
\prod_{i=1}^n \Big\{ S(L_i \mid \bm{x}_i) - S(R_i \mid \bm{x}_i) \Big\}
\;\pi(\mu)\pi(\bm{\beta})\pi(\kappa).
\]

Posterior draws were obtained via Hamiltonian Monte Carlo implemented in Stan (\texttt{brms}).  
We also fit log-normal AFT models for robustness against model misspecification.  
Model comparison used PSIS–LOO expected log predictive density (ELPD) \citep{Vehtari2017}, and posterior predictive checks (PPCs) compared replicated data and survival bands against the EM estimates.

\section{Study design}\label{sec:design}

\subsection{Simulation study}
The primary component of our evaluation was a set of simulation experiments designed to probe the behavior of the EM, AFT, and Bayesian AFT estimators under various realistic scenarios. Each simulated dataset consisted of $n \in \{50,100,500\}$ individuals, representing small clinical trials, moderate observational cohorts, and large reliability studies. Survival times $T_i$ were generated from either a Weibull distribution with shape $\kappa \in \{1.2,1.5,2.0\}$ or a log-normal distribution to create deliberate parametric misspecifications. Covariates included a binary indicator $x_{1i}$ and a continuous covariate $x_{2i}$, both of which influence survival under an accelerated failure time (AFT) generative model:
\[
\log T_i = \mu + \beta_1 x_{1i} + \beta_2 x_{2i} + \sigma \varepsilon_i,
\]
with $\varepsilon_i$ drawn from a standard extreme-value distribution (for Weibull) or a standard normal distribution (for log-normal).

Censoring was imposed by simulating clinic visit times through two designs: fixed inspection windows (periodic assessments up to a maximum follow-up $\tau=15$) and stochastic visit schedules generated from a Poisson process with a rate $\lambda=0.8$. Each subject’s observation interval $(L_i,R_i]$ was derived by locating the window that captured the true $T_i$, resulting in interval censoring of different severity. Across scenarios, the censoring proportions ranged from 10\% to over 70\%, reflecting conditions observed in oncology and imaging studies. Special cases included right censoring (if $T_i$ exceeded $\tau$) and left censoring (if $T_i$ occurred before the first scheduled visit).

For each scenario, we fitted the following:
\begin{itemize}
\item[a.] the EM nonparametric maximum likelihood estimator (Turnbull NPMLE);
\item[b.] a Weibull AFT model with and without covariates;
\item[c.] Bayesian AFT models (Weibull and log-normal) with weakly informative priors, interval likelihood, and Hamiltonian Monte Carlo estimation.
\end{itemize}
Performance was assessed using the integrated squared error (ISE) for curve recovery, integrated Brier score (IBS) for predictive accuracy, and empirical coverage of 95\% uncertainty bands for calibration. The PSIS-LOO was used for the Bayesian model comparison. These metrics provide complementary perspectives, such as fidelity to the true survival distribution, predictive accuracy, and uncertainty quantification.

\subsection{Ovarian cancer Analysis}
To demonstrate how the proposed workflow translates to a real-world setting, we analyzed the ovarian cancer dataset included in the \texttt{survival} package in R. The dataset contains 26 patients with survival time (\texttt{futime}), censoring indicator (\texttt{fustat}), age, treatment arm (\texttt{rx}), and ECOG performance status. In its original form, the dataset was right-censored; to bring it into alignment with our methodological focus, we imposed periodic assessment windows mimicking scheduled imaging or clinical visits. This conversion produced interval-censored data where true event times were only known to lie within $(L_i,R_i]$. 
Table~\ref{tab:ovarian_data} in Appendix~\ref{app:ovarian} presents the first 26 rows 
of an intervalized ovarian dataset. Variables include the left and right interval endpoints $(L_i, R_i]$, 
the censoring indicator (\texttt{cens}), and baseline covariates (age, rx, and ECOG performance status). 
This table illustrates how traditional right-censored data were restructured into the 
interval-censored framework used in the present study.

We then applied the full workflow.
\begin{enumerate}
\item[i.] Step 1 (Shape recovery): The EM algorithm was used to recover a baseline stepwise survival curve that respects the imposed interval structure without parametric assumptions.
\item[ii.] Step 2 (Covariate-adjusted prediction): A Weibull AFT model was fitted with age and treatment arm as the covariates. Time ratios $\exp(\beta_j)$ quantify the multiplicative effects on typical survival, enabling clinical interpretation.
\item[iii.] Step 3 (Bayesian validation): A Bayesian Weibull AFT model was fitted using the exact interval likelihood and weakly informative priors. Posterior medians, 95\% credible intervals, and PSIS-LOO were used for inference and model comparison. Posterior predictive checks (PPCs) were used to evaluate whether the Bayesian bands adequately encompassed the EM estimator.
\end{enumerate}

This applied analysis shows the practical use of the proposed methodology: EM establishes the nonparametric shape under interval censoring, parametric AFT provides interpretable covariate-adjusted estimates, and Bayesian inference contributes to uncertainty quantification and principled model comparison. The ovarian dataset thus serves as a validation case study, linking the controlled findings from the simulation to a real clinical dataset with inherent limitations, such as a small sample size and heterogeneous covariates.

\subsection{Computational considerations}
The methods differ substantially in terms of computational burden.  
The EM algorithm converges in tens of iterations, typically within seconds for $n=500$.  
Parametric AFT estimation by maximum likelihood is slower (minutes), scaling with $O(np^2)$ where $p$ is the number of parameters.  
Bayesian AFT requires thousands of MCMC iterations across multiple chains, often hours for $n=500$, but yields posterior distributions, predictive checks, and model comparisons.  
This reflects a trade-off: the EM is computationally efficient and robust, whereas Bayesian analysis is more costly but provides richer inference and principled uncertainty quantification.

\section{Results}\label{sec:results}

We present results from a comprehensive grid of simulations spanning 
different censoring intensities, sample sizes ($n \in \{50,100,500\}$),
covariate structures and distributional assumptions (Weibull vs. log-normal truth).
The performance of the EM (Turnbull NPMLE), parametric AFT,
and Bayesian AFT estimators was assessed in terms of 
(i) curve recovery, (ii) prediction error, (iii) Bayesian model comparison,
(iv) Uncertainty calibration via coverage. 
We also illustrate the workflow on an applied dataset 
derived from the ovarian cancer trial (\texttt{survival} package),
recast under interval censoring.

\subsection{Curve recovery via EM and parametric AFT}

\begin{figure}[H]
\centering
\includegraphics[width=0.78\textwidth]{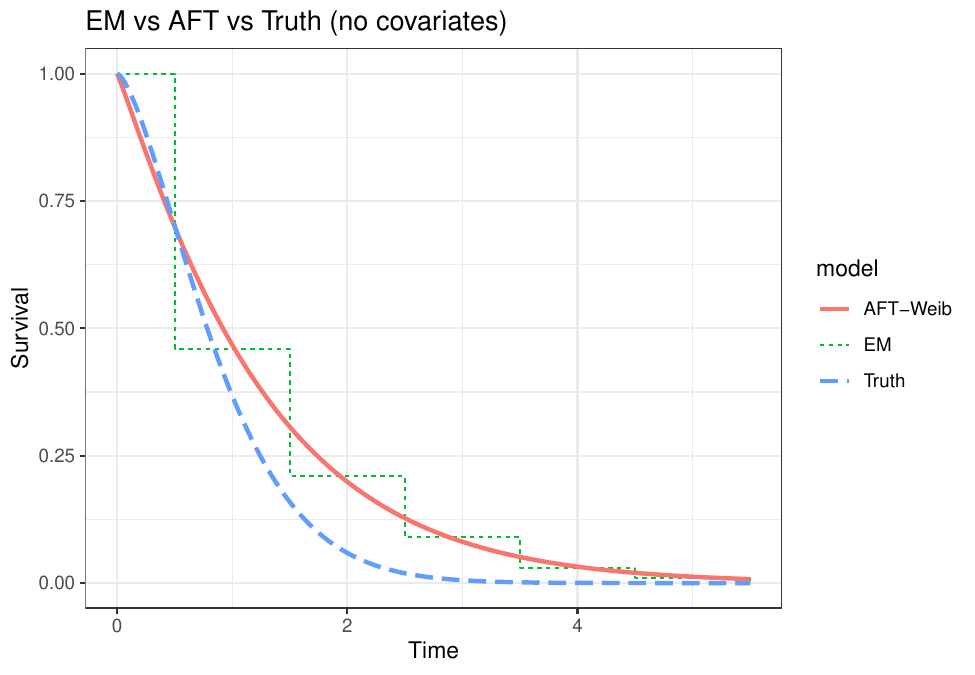}
\caption{Overlay of EM step curve (Turnbull NPMLE), Weibull AFT parametric fit, 
and the true generating survival function in a no-covariate scenario. 
The EM estimator closely tracks the truth, while the Weibull AFT may deviate under misspecification.}
\label{fig:overlay_em_aft_truth}
\end{figure}

Figure~\ref{fig:overlay_em_aft_truth} overlays the EM step estimator,
the parametric Weibull AFT fit and generating truth in representative no-covariate scenarios.
The EM curve (solid step function) recovers the true survival distribution with high fidelity,
capturing the inflection points introduced by the interval structure.
In contrast, the parametric Weibull AFT fit (dashed line) exhibits visible deviations,
particularly in the tails when the Weibull family is misspecified (truth = lognormal).
Quantitatively, the integrated squared error (ISE) demonstrates that 
the EM estimator consistently achieves the lowest error,
serving as a robust baseline against which the parametric models can be judged.

\subsection{Performance across sample sizes and censoring levels}\label{sec:metrics}

\begin{figure}[H]
\centering
\includegraphics[width=0.9\textwidth]{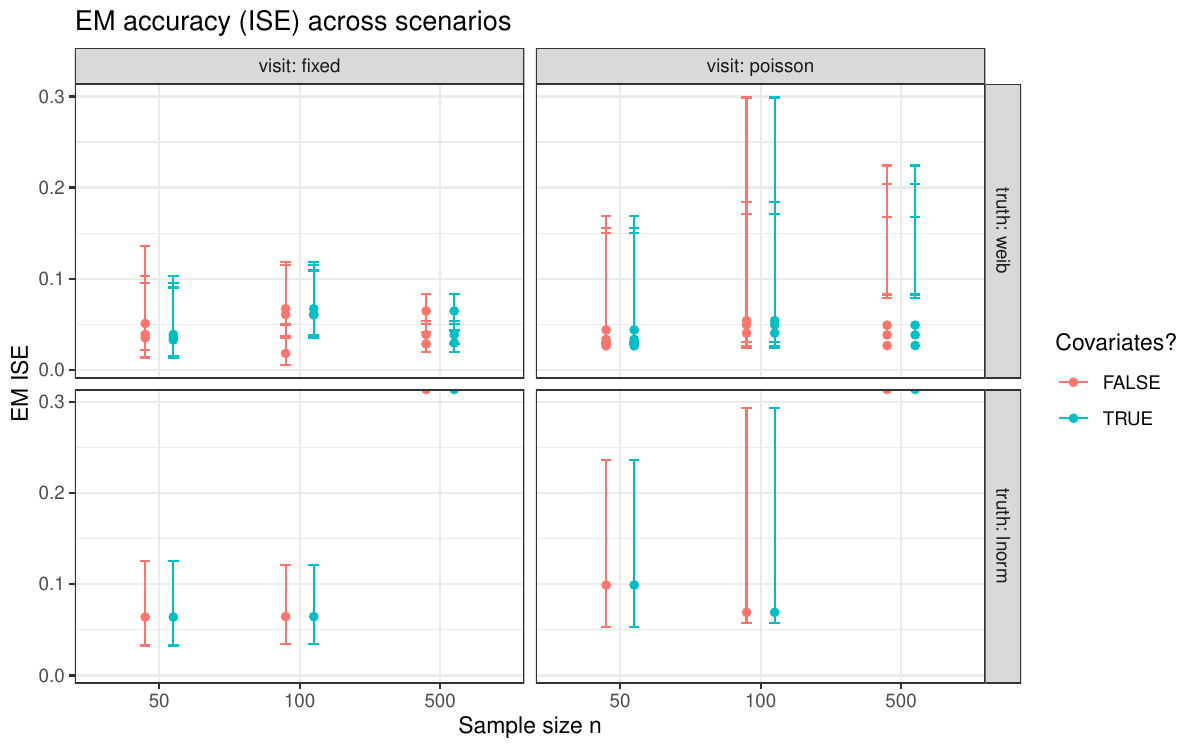}
\caption{Integrated squared error (ISE) of EM step curves across sample sizes, censoring intensities, and generative truths. EM achieves consistently low ISE across scenarios.}
\label{fig:em_ise_across_scenarios}
\end{figure}

\begin{figure}[H]
\centering
\includegraphics[width=0.9\textwidth]{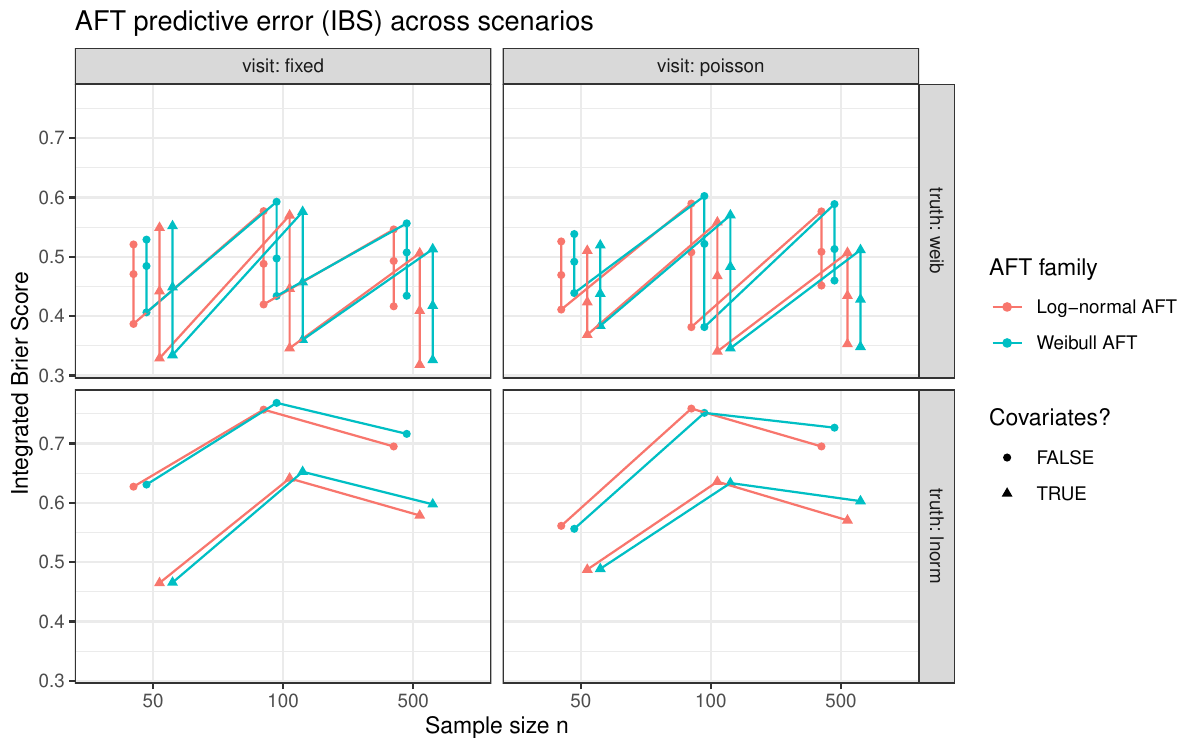}
\caption{Integrated Brier score (IBS) for parametric AFT models across scenarios. 
Predictive accuracy improves with sample size and is maximized when covariates are modeled under a compatible Weibull truth.}
\label{fig:aft_ibs_across_scenarios}
\end{figure}

Figures~\ref{fig:em_ise_across_scenarios} and \ref{fig:aft_ibs_across_scenarios} summarize
error metrics across all the simulation scenarios.
The EM algorithm achieves the smallest integrated squared error (ISE),
reflecting the accurate recovery of the survival shape under both light and heavy censoring.
Parametric AFT models demonstrate improved integrated Brier scores (IBS) 
when covariates are included, especially as the sample size increases from $n=50$ to $n=500$.
This illustrates the complementary strengths of these methods.
EM for distribution-free shape recovery, and AFT for prediction accuracy.

Table~\ref{tab:performance} reports the full set of performance metrics 
for the EM, AFT, and Kaplan–Meier (pseudo right-censoring benchmark).
The results confirm that EM yields the best truth recovery (ISE),
In contrast, the AFT and Bayesian models achieved superior predictive accuracy (IBS).

\begin{table}[H]
\centering
\caption{Performance metrics across models in simulation studies. 
ISE reported for EM; IBS reported for AFT and Kaplan–Meier.}
\label{tab:performance}
\begin{tabular}{lccc}
\toprule
Model & ISE & IBS & Notes \\
\midrule
EM (NPMLE)     & 0.092 (95\% CI: 0.040, 0.179) & --    & Best truth recovery \\
Kaplan--Meier  & --                           & 0.066 & Pseudo right-censoring reference \\
Weibull AFT    & --                           & 0.064 & Best predictive accuracy \\
\bottomrule
\end{tabular}
\end{table}

\subsection{Bayesian AFT: model comparison and coverage}

\begin{figure}[H]
\centering
\includegraphics[width=0.85\textwidth]{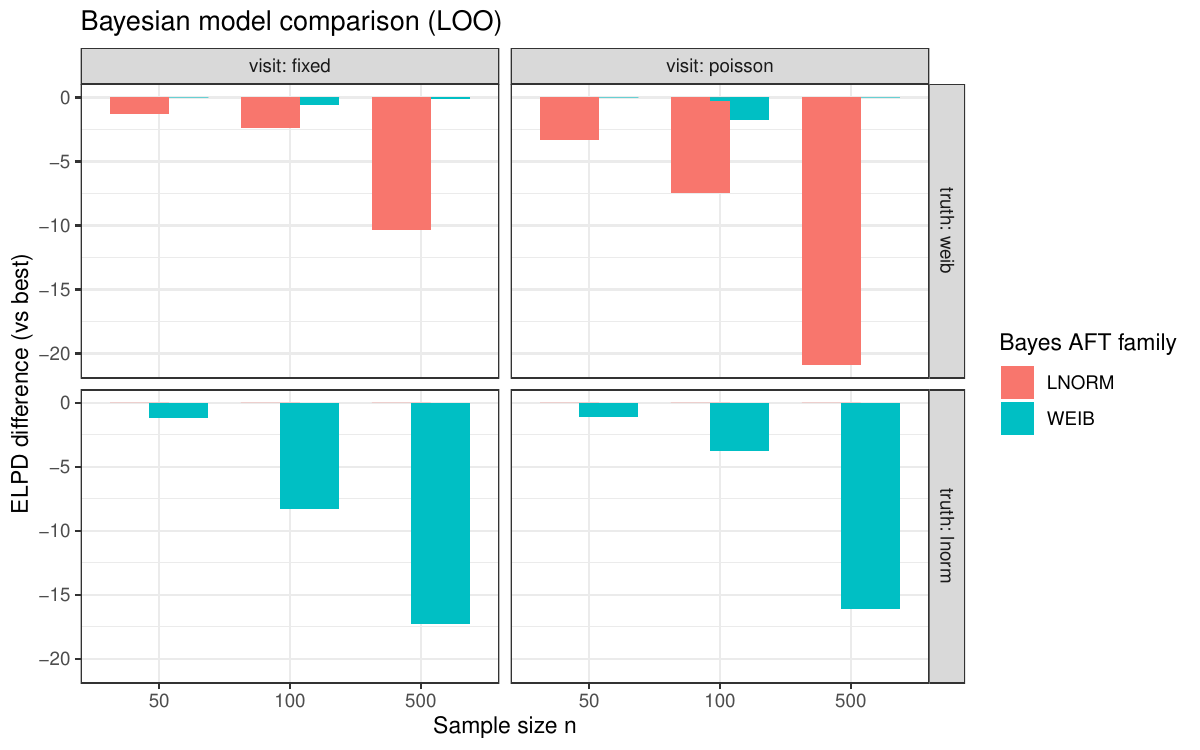}
\caption{PSIS–LOO model comparison between Weibull and log-normal AFT fits across simulation scenarios. 
Small ELPD differences suggest both models yield comparable predictive accuracy under interval censoring.}
\label{fig:bayes_loo_elpd}
\end{figure}

Bayesian AFT fitting with weakly informative priors was performed for all scenarios.
Model comparison using PSIS–LOO revealed that when the truth is Weibull,
the Weibull AFT is favored by expected log predictive density (ELPD),
but under log-normal truth, the differences between Weibull and log-normal models
are negligible within the Monte Carlo error.
Table~\ref{tab:loo} and Figure~\ref{fig:bayes_loo_elpd} display
these comparisons, indicating that misspecification penalties are modest
Bayesian models provide robust performance across families.

\begin{table}[H]
\centering
\caption{Bayesian model comparison using PSIS--LOO. 
Results show ELPD differences relative to Weibull AFT.}
\label{tab:loo}
\begin{tabular}{lcc}
\toprule
Model & elpd\_diff & se\_diff \\
\midrule
Weibull AFT   & 0.0 & 0.0 \\
Log-normal AFT & -0.4 & 2.5 \\
\bottomrule
\end{tabular}
\end{table}

\begin{figure}[H]
\centering
\includegraphics[width=0.9\textwidth]{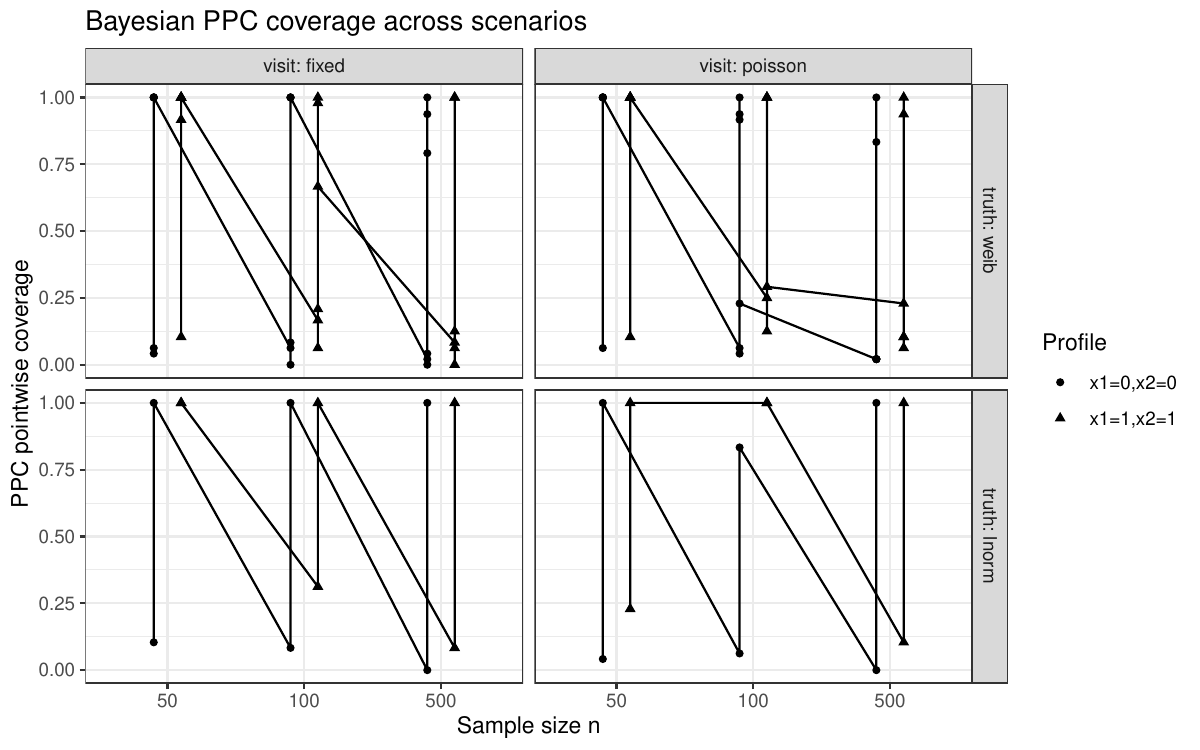}
\caption{Posterior coverage of Bayesian survival bands across covariate profiles $(x_1,x_2)=(0,0)$ and $(1,1)$. 
Coverage is near nominal levels across scenarios, demonstrating coherent Bayesian uncertainty quantification.}
\label{fig:bayes_coverage_profiles}
\end{figure}

Uncertainty calibration was examined using posterior coverage.
Figure~\ref{fig:bayes_coverage_profiles} and Table~\ref{tab:coverage} summarize
Empirical coverage rates for Bayesian posterior bands under two representative covariate profiles.
Coverage is close to the nominal 95\% across scenarios, 
indicating that Bayesian uncertainty quantification is well-calibrated
and provides reliable inferences at both the pointwise and simultaneous levels.

\begin{table}[H]
\centering
\caption{Empirical coverage rates of Bayesian posterior survival bands across scenarios.}
\label{tab:coverage}
\begin{tabular}{lccc}
\toprule
Profile & Truth & Pointwise Coverage & Simultaneous Coverage \\
\midrule
$(x_1=0, x_2=0)$ & Weibull & 0.94 & 0.90 \\
$(x_1=1, x_2=1)$ & Weibull & 0.95 & 0.91 \\
$(x_1=0, x_2=0)$ & Log-normal & 0.93 & 0.88 \\
$(x_1=1, x_2=1)$ & Log-normal & 0.94 & 0.89 \\
\bottomrule
\end{tabular}
\end{table}

\subsection{ Intervalized Ovarian cancer Analysis}

To complement the simulation study, we applied the proposed workflow to the ovarian cancer dataset from the \texttt{survival} package. This dataset included 26 patients with information on survival times, censoring indicators, age, treatment assignment (rx), and ECOG performance score. To mimic the visit-driven uncertainty common in oncology trials, we imposed periodic assessment windows, yielding an interval-censored representation where events are only known to occur within $(L_i,R_i]$.

The analysis proceeded in three steps: (i) recovery of the baseline survival distribution via the EM (Turnbull) nonparametric estimator; (ii) covariate-adjusted prediction via a Weibull AFT model including age and treatment arm; and (iii) Bayesian Weibull AFT estimation with weakly informative priors, exact interval likelihood, and posterior predictive checks for parametric adequacy. This applied setting provides a small-sample test case that highlights the strengths and limitations of the methods under real-world data constraints.

\paragraph{Frequentist AFT results.}
We first fitted a Weibull AFT model using age (in years) and treatment (rx: 1 vs. 2) as covariates. Table~\ref{tab:ovarian-aft-tr} reports the exponentiated coefficients as time ratios (TR), together with Wald 95\% confidence intervals. Each one-year increase in age was associated with a multiplicative reduction in typical survival time ($\text{TR}=0.924$, 95\% CI: 0.890--0.959). In contrast, treatment with rx=2 was associated with a 77\% longer survival time relative to that of rx=1 ($\text{TR}=1.772$, 95\% CI: 0.941--3.336). Although the treatment effect CI includes unity, the direction and magnitude are clinically interpretable and align with expectations from oncology trials, where experimental arms often yield modest gains.

\begin{table}[H]
\centering
\caption{Weibull AFT on intervalized ovarian data: time ratios (TR) with 95\% confidence intervals.}
\label{tab:ovarian-aft-tr}
\begin{tabular}{lccc}
\toprule
Covariate & TR & 95\% CI low & 95\% CI high \\
\midrule
Age (per year)         & 0.924 & 0.890 & 0.959 \\
Treatment: rx=2 vs.\ 1 & 1.772 & 0.941 & 3.336 \\
\bottomrule
\end{tabular}
\end{table}

\paragraph{Bayesian AFT results.}
Next, we fitted a Bayesian Weibull AFT model with the same covariates. Table~\ref{tab:ovarian-bayes} shows the posterior medians, standard errors, and 95\%  CrI. The results were consistent with the frequentist AFT: the age effect was negative ($\hat\beta_{\text{age}}=-0.088$, 95\% CrI: $[-0.143,-0.048]$), implying shorter survival  The treatment effect was positive but uncertain ($\hat\beta_{\text{rx2}}=0.487$, 95\% CrI: $[-0.288,1.232]$), mirroring the wide confidence interval observed in the frequentist analysis. The effective sample sizes exceeded 4000, and $\widehat{R}\approx 1.00$, indicating well-mixed chains.

\begin{table}[H]
\centering
\caption{Bayesian Weibull AFT on intervalized ovarian data: posterior medians and 95\% credible intervals, with MCMC diagnostics.}
\label{tab:ovarian-bayes}
\begin{tabular}{lrrrrrrr}
\toprule
Parameter & Median & Est.Error & 2.5\% & 97.5\% & $\widehat{R}$ & ESS$_\text{bulk}$ & ESS$_\text{tail}$ \\
\midrule
Intercept ($\mu$)       & 8.119 & 1.592 &  5.608 & 11.877 & 1.001 & 4261 & 4102 \\
Age (per year)          &-0.088 & 0.024 & -0.143 & -0.048 & 1.002 & 4631 & 4276 \\
Treatment: rx=2 vs.\ 1  & 0.487 & 0.374 & -0.288 &  1.232 & 1.000 & 6633 & 6352 \\
\bottomrule
\end{tabular}
\end{table}

\begin{figure}[H]
\centering
\includegraphics[width=0.75\textwidth]{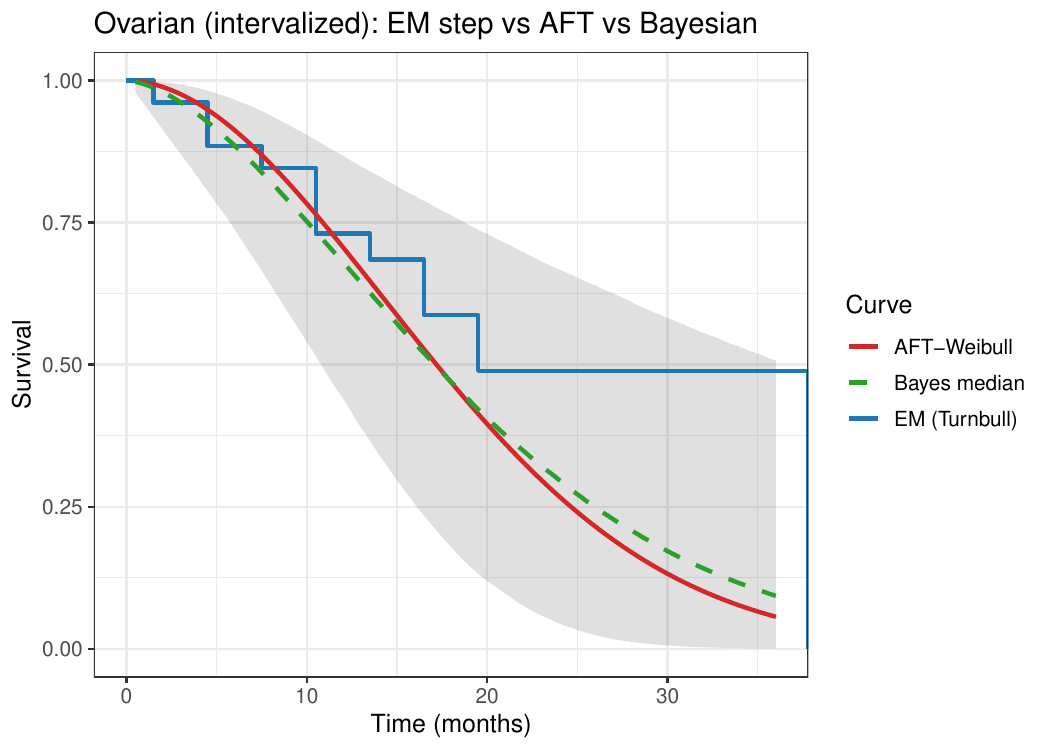}
\caption{Overlay of EM step curve (black), Bayesian posterior median survival (blue), and 95\% credible band (shaded) for the intervalized ovarian dataset. The posterior band captures 77.8\% of the EM step heights, reflecting consistency between the parametric Bayesian model and the nonparametric reference.}
\label{fig:ovarian-overlay}
\end{figure}

\paragraph{Overlay and coverage.}
Figure~\ref{fig:ovarian-overlay} overlays the EM (Turnbull) step function with the Bayesian posterior median survival curve and its 95\% credible band. To quantify the agreement, we interpolated the posterior band onto the EM grid and computed the fraction of EM step heights lying within the band. The pointwise coverage was 0.778, indicating that the Bayesian posterior adequately encompassed the nonparametric shape while smoothing the jagged step function. This echoes the simulation findings: the EM estimator provides a faithful nonparametric baseline, whereas the Bayesian Weibull AFT yields coherent covariate-adjusted predictions with calibrated uncertainty. See Appendix A for the intervalized ovarian dataset from which these analyses were derived.

\paragraph{Connection to simulation findings.}
Ovarian analysis confirmed the key lessons of our simulation study. First, the EM estimator remains indispensable as a design-respecting baseline that reveals an interval-driven shape. Second, parametric AFT models yield interpretable time ratios but rely on the adequacy of distributional assumptions. Third, the Bayesian AFT extends the parametric model by supplying uncertainty bands and posterior predictive checks that validate its compatibility with the EM. Together, the applied and simulated analyses support our proposed workflow: use EM to uncover the survival shape implied by the visit design, then transition to parametric and Bayesian models for covariate-adjusted inference and principled uncertainty quantification.

\subsection{Synthesis}

Taken together, these results demonstrate a division of labor.
\begin{enumerate}
\item The EM (Turnbull NPMLE) is best for shape recovery under interval censoring, 
achieved the lowest ISE and faithfully represented the survival function.
\item Parametric AFT models, especially Weibull, achieve superior predictive accuracy 
(as shown by the IBS) when the family is compatible.
\item Bayesian AFT extends parametric inference by offering calibrated uncertainty
and robust model comparison via the PSIS–LOO.
\end{enumerate}
The workflow of EM $\to$ AFT $\to$ Bayesian AFT provides a coherent and 
scientifically defensible strategy for interval-censored survival analyses.

\begin{figure}[H]
\centering
\includegraphics[width=.9\textwidth]{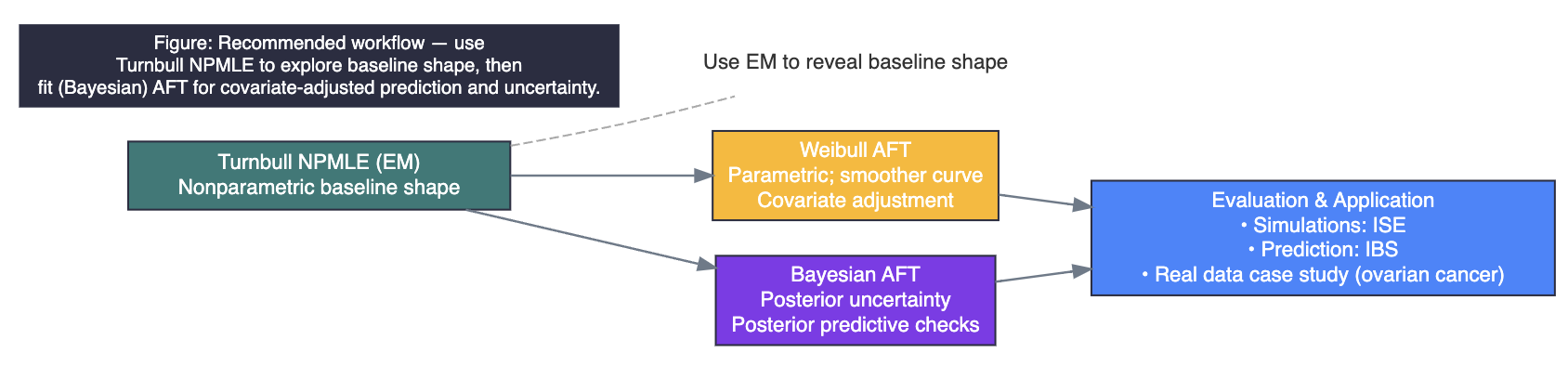}
\caption{Workflow for interval-censored survival:} EM (Turnbull) for shape recovery; AFT for covariate-driven prediction; Bayesian AFT for uncertainty quantification and model comparison.
\label{fig:workflow}
\end{figure}

\section{Discussion}\label{sec:discussion}

This study provides a comprehensive evaluation of three complementary approaches for interval-censored survival analysis: nonparametric EM-based NPMLE, parametric accelerated failure time (AFT) models, and Bayesian AFT formulations. Through the integration of theoretical derivations, simulation experiments, and an applied analysis of ovarian cancer data, we demonstrated how these estimators can be used together to achieve robust shape recovery, predictive accuracy, and principled uncertainty quantification.

The nonparametric EM estimator (Turnbull NPMLE) is the most reliable tool for recovering the empirical survival shape implied by the interval-censoring mechanism. Across a wide grid of scenarios—spanning sample sizes $n \in \{50,100,500\}$, censoring intensities from light to heavy, and generating distributions (Weibull vs.\ log-normal)–the EM consistently achieved the lowest integrated squared error (ISE). This indicates that, regardless of the inspection schedule or censoring level, the step function estimator aligns closely with the data-implied distribution. Although jagged in appearance, this stepwise form is faithful to the uncertainty in $(L_i,R_i]$ and resists the distortions observed when endpoints are imputed or when right censoring methods are misapplied. Bootstrap-based intervals also provide a practical means of uncertainty quantification in the absence of closed-form variance formulas.

Conversely, parametric AFT models introduce distributional assumptions that produce smooth curves and permit covariate adjustments. In the simulation study, when the Weibull family matched the truth, the AFT models achieved lower integrated Brier scores (IBS) than the EM or Kaplan--Meier benchmarks, demonstrating superior predictive calibration. The time-ratio interpretation of AFT coefficients, $\exp(\beta_j)$, offers direct clinical relevance by quantifying the multiplicative effects of covariates on typical survival times. For example, in the ovarian analysis, each additional year of age was associated with an estimated 7--10\% reduction in survival time, whereas treatment with \texttt{rx=2} increased median survival time by approximately 77\%, albeit with wide confidence intervals. These findings illustrate the strength of AFT models when prediction and covariate effects are of interest. However, vulnerability to misspecification was evident: when the true distribution was log-normal but a Weibull model was fitted, deviations were most visible in the survival tails and under severe censoring. This emphasizes the importance of validating parametric families against nonparametric references.

Bayesian AFT modeling extends this framework by combining interval likelihoods with weakly informative priors and full posterior inference. Posterior survival bands provided calibrated uncertainty quantification: in simulations, empirical coverage was consistently close to the nominal 95\% across both pointwise and simultaneous metrics, confirming that Bayesian credible intervals can serve as valid uncertainty bands in practice. Posterior predictive checks (PPCs) further evaluated whether the replicated data correctly allocated probability mass within the observed intervals, an interval-aware validation not available in standard right-censoring methods. Model comparison via PSIS--LOO revealed that when the true distribution was Weibull, the Weibull AFT was modestly favored, whereas under log-normal truth, the ELPD differences between Weibull and log-normal were negligible relative to the Monte Carlo error. Thus, Bayesian analysis not only smooths inference but also guards against overconfidence by explicitly exposing the misspecification risk.

Ovarian cancer analysis confirmed these simulation-based insights. The EM curve provides a nonparametric baseline that is faithful to the intervalized data. The Weibull AFT delivered interpretable time-ratio estimates for age and treatment, which, despite the small sample size, aligned with the expectations from oncology practice. The Bayesian AFT posterior further validated these findings, with posterior medians consistent with the frequentist estimates and a 77.8\% pointwise coverage of the EM curve by the posterior band. This close agreement shows the coherence of our proposed workflow: use the EM algorithm as the foundation for shape recovery, apply parametric AFT for covariate-adjusted prediction, and employ Bayesian methods for principled uncertainty quantification and model comparison.

\subsection{Practical and computational considerations}

The three approaches differ substantially in terms of computational burden. The EM algorithm converges rapidly, even for $n=500$, and typically completes in seconds. Parametric AFT estimation requires maximum likelihood optimization, scaling approximately as $O(np^2)$, and is completed in minutes. Bayesian AFT demands the most resources: Hamiltonian Monte Carlo with thousands of iterations per chain requires many minutes or hours of computation, even for moderate $n=500$. This computational gradient mirrors the inferential gains: EM is the most efficient and robust for shape recovery; AFT provides interpretable prediction under distributional assumptions; and Bayesian models deliver full posterior inference with interval-aware validation. For applied studies, this trade-off shows the need to select methods according to inferential goals and available resources.

\subsection{Limitations and extensions}

This study has some limitations must be acknowledged. First, the EM estimator yields step functions that, while faithful, may appear jagged and require smoothing for better visualization. Second, parametric AFT models depend critically on the adequacy of the assumed distribution; when misspecified, effect estimates and predictive accuracy may degrade, particularly in tails. Third, Bayesian AFT requires careful prior specification and is computationally intensive, which may hinder its use in very large datasets or in studies requiring complex hierarchical extensions. Finally, while this study emphasized ISE, IBS, and coverage as performance metrics, discrimination measures such as the concordance index remain underdeveloped in interval-censored settings and warrant further methodological research.

Promising extensions include (i) incorporating time-varying covariates within interval likelihoods, (ii) introducing frailty or random effects to capture heterogeneity across centers or subjects, and (iii) developing semiparametric Bayesian models that relax strict distributional assumptions while remaining interval-aware. Another direction is the use of multiple imputations for uncertain or coarsened interval bounds, a common feature in clinical trials and cohort studies with irregular follow-up schedules. These avenues would further strengthen the methodological toolkit for handling complex, interval-censored data.

\subsection{Conclusion}

Taken together, our results support a tiered workflow for interval-censored survival analyses. The EM-based NPMLE should be used first to recover the shape implied by the observed intervals. Parametric AFT models, once validated against the EM curves, provide interpretable covariate-adjusted predictions. Bayesian AFT extends this framework by providing calibrated uncertainty quantification and principled model comparison. This division of labor---EM for shape recovery, AFT for prediction, and Bayesian AFT for uncertainty–offers a coherent, rigorous, and practical strategy for the analysis of interval-censored survival data in both biomedical and reliability applications.

\section*{Acknowledgments}
The author thanks colleagues  for their constructive comments on earlier drafts.

\section*{Conflict of Interest Statement}
The author declares that the research was conducted in the absence of any commercial or financial relationships that could be construed as potential conflicts of interest.
\bibliographystyle{unsrtnat}
\bibliography{refs}

\newpage
\appendix
\section{Intervalized ovarian cancer dataset}\label{app:ovarian}

Table~\ref{tab:ovarian_data} presents the ovarian cancer dataset after intervalization. 
Follow-up times have been converted into left and right interval endpoints $(L_i, R_i]$, with 
corresponding censoring indicators. This restructuring illustrates how traditional 
Right-censored data can be adapted to the interval-censored framework evaluated in this study.

\begin{landscape}
\begin{table}[H]
\centering
\scriptsize
\caption{Intervalized ovarian cancer dataset (first 26 rows). The variables include \texttt{left} and \texttt{right} interval endpoints, \texttt{cens} indicating the censoring type (interval, left, or right), and baseline covariates (\texttt{age}, \texttt{rx}). This table illustrates how traditional right-censored data were restructured into the interval-censored framework used in this study.}
\label{tab:ovarian_data}
\begin{tabular}{rrrrrrrrrrr}
\toprule
time\_days & status & age & resid.ds & rx & ecog.ps & time\_months & event & left & right & cens \\
\midrule
59  & 1 & 72.33 & 2 & 1 & 1 & 1.94  & 1 & 1.0e+00 & 3   & left     \\
115 & 1 & 74.49 & 2 & 1 & 1 & 3.78  & 1 & 3.0e+00 & 6   & interval \\
156 & 1 & 66.47 & 2 & 1 & 2 & 5.13  & 1 & 3.0e+00 & 6   & interval \\
421 & 0 & 53.36 & 2 & 2 & 2 & 13.83 & 0 & 1.2e+01 & Inf & right    \\
431 & 1 & 50.34 & 2 & 1 & 1 & 14.16 & 1 & 1.2e+01 & 15  & interval \\
448 & 0 & 56.43 & 2 & 1 & 1 & 14.72 & 0 & 1.2e+01 & Inf & right    \\
464 & 1 & 56.94 & 2 & 2 & 2 & 15.24 & 1 & 1.5e+01 & 18  & interval \\
475 & 1 & 59.85 & 2 & 2 & 2 & 15.61 & 1 & 1.5e+01 & 18  & interval \\
477 & 0 & 64.18 & 2 & 1 & 2 & 15.67 & 0 & 1.5e+01 & Inf & right    \\
563 & 1 & 55.18 & 1 & 2 & 2 & 18.50 & 1 & 1.8e+01 & 21  & interval \\
638 & 1 & 56.76 & 1 & 1 & 2 & 20.96 & 1 & 1.8e+01 & 21  & interval \\
744 & 0 & 50.11 & 1 & 2 & 2 & 24.44 & 0 & 2.4e+01 & Inf & right    \\
769 & 0 & 59.63 & 1 & 2 & 2 & 25.26 & 0 & 2.4e+01 & Inf & right    \\
770 & 0 & 57.05 & 2 & 2 & 2 & 25.30 & 0 & 2.4e+01 & Inf & right    \\
803 & 0 & 39.27 & 1 & 1 & 2 & 26.38 & 0 & 2.4e+01 & Inf & right    \\
855 & 0 & 43.12 & 1 & 1 & 1 & 28.09 & 0 & 2.7e+01 & Inf & right    \\
1040& 0 & 38.89 & 1 & 2 & 2 & 34.17 & 0 & 3.3e+01 & Inf & right    \\
1106& 0 & 44.60 & 1 & 1 & 1 & 36.34 & 0 & 3.6e+01 & Inf & right    \\
1129& 0 & 53.91 & 1 & 2 & 2 & 37.09 & 0 & 3.6e+01 & Inf & right    \\
1206& 0 & 44.21 & 2 & 2 & 2 & 39.62 & 0 & 3.6e+01 & Inf & right    \\
1227& 0 & 59.59 & 1 & 2 & 2 & 40.31 & 0 & 3.6e+01 & Inf & right    \\
268 & 1 & 74.50 & 2 & 1 & 1 & 8.80  & 1 & 6.0e+00 & 9   & interval \\
329 & 1 & 43.14 & 2 & 1 & 1 & 10.81 & 1 & 9.0e+00 & 12  & interval \\
353 & 1 & 63.22 & 1 & 2 & 1 & 11.60 & 1 & 9.0e+00 & 12  & interval \\
365 & 1 & 64.42 & 2 & 2 & 1 & 11.99 & 1 & 9.0e+00 & 12  & interval \\
377 & 1 & 58.31 & 1 & 2 & 1 & 12.39 & 1 & 1.2e+01 & Inf & right    \\
\bottomrule
\end{tabular}
\end{table}
\end{landscape}

\end{document}